\documentclass[conference]{IEEEtran}
\usepackage{graphicx} 

\IEEEoverridecommandlockouts
\usepackage{cite}
\usepackage{amsmath,amssymb,amsfonts}
\usepackage{algorithmic}
\usepackage{graphicx}
\usepackage{textcomp}
\usepackage{xcolor}
\usepackage{xurl}
\usepackage{float}
\usepackage{optidef}
\def\BibTeX{\rm B\kern-.05em{\sc i\kern-.025em b}\kern-.08em
    T\kern-.1667em\lower.7ex\hbox{E}\kern-.125emX}
\begin{document}

\title{Open-Loop and Model Predictive Control for Electric Vehicle Charging to Manage Excess Renewable Energy Supply in Texas}

\author{
\IEEEauthorblockN{Kelsey M. Nelson\IEEEauthorrefmark{1}, Maureen S. Golan\IEEEauthorrefmark{1}, Matthew D. Bartos\IEEEauthorrefmark{1}, Javad Mohammadi\IEEEauthorrefmark{1}}
\IEEEauthorblockA{\IEEEauthorrefmark{1}{Maseeh Department of Civil, Architectural and Environmental Engineering} \\
{The University of Texas at Austin}\\
Austin, USA \\\
{\{kelseynelson, mgolan, mdbartos, javadm}\}@utexas.edu}

}

\maketitle
\begin{abstract} 

Modern power grids are evolving to become more interconnected, include more electric vehicles (EVs), and utilize more renewable energy sources (RES). Increased interconnectivity provides an opportunity to manage EVs and RES by using price signaling to shift EV loads towards periods of high RES output. This work uses ERCOT's 2035 RES installation plans and projections for Texas's EV fleet to examine and compare how both open-loop control and model predictive control (MPC) schemes can leverage time varying rates for EV charging to utilize excess RES supply that may otherwise be underutilized in a highly weather-dependent grid. The results show that while open-loop control increases RES usage, MPC increases RES usage even further by responding to RES outputs that differ from forecasts due to the inherent uncertainty of weather predictions. If MPC is used with time steps that are too frequent, however, difficulties arise; EV owners may find it too onerous to keep up with changing price structures, and frequent over-corrections to charging profiles can lead to a ``rebound peak" phenomenon. Therefore, control schemes should balance maximizing RES usage with ensuring customer participation. 

\begin{IEEEkeywords}
Electric Vehicles, Electrical Grid Impact, GHG emissions, Model Predictive Control
\end{IEEEkeywords}

\end{abstract}

\section*{Nomenclature}
\begin{table}[htbp]
\caption{2035 parameters for MPC model inputs} 
\label{Table1}
\begin{center}
\begin{tabular}{|c|p{5.5cm}|}
\hline
$BAU_{dem}$ & The business as usual (uncontrolled) demand of the EV fleet at time k\\
\hline
$EVt$ & The total number of EVs on the road \\
\hline
$k$ & The kth hour within the optimizer\\
\hline
$k_h$ & The time horizon of the optimizer\\
\hline
$L$ & The aggregated, controlled EV load as a result of the control scheme being implemented\\
\hline
$LDVt$ & The total number of light duty vehicles on the road\\
\hline
$MS$ & The EV market share, or percentage of light duty vehicles sold in a year that are electric\\
\hline
$n$ & The year for eq. \eqref{EV_total_N}\\
\hline
$P_{e}$ & The percentage of BAU load that a price signal is able to shift to the next time step\\
\hline
$R$ & The average lifetime of a light duty vehicle\\
\hline
$RES_e$ & The excess RES profile for the optimizer\\
\hline

 \end{tabular}
 \end{center}
\end{table}

\section{Introduction}
\subsection{Background and Motivation}
\label{Background and Motivation}

Electric vehicles are becoming more common in the United States as the nation's administration incentivizes their adoption and more consumers become aware of their environmental benefits \cite{Registrations}. As of 2024, there are about 300,000 EVs on the roads in Texas \cite{Registrations}, but if current EV adoption goals are met (50\% market share by 2030), this number could rise to nearly 10 million EVs on Texas' roads by 2035 \cite{EVTarget} \cite{Keith_2019}. 

\begin{figure}[htbp]
\centering
\includegraphics[scale=.8]{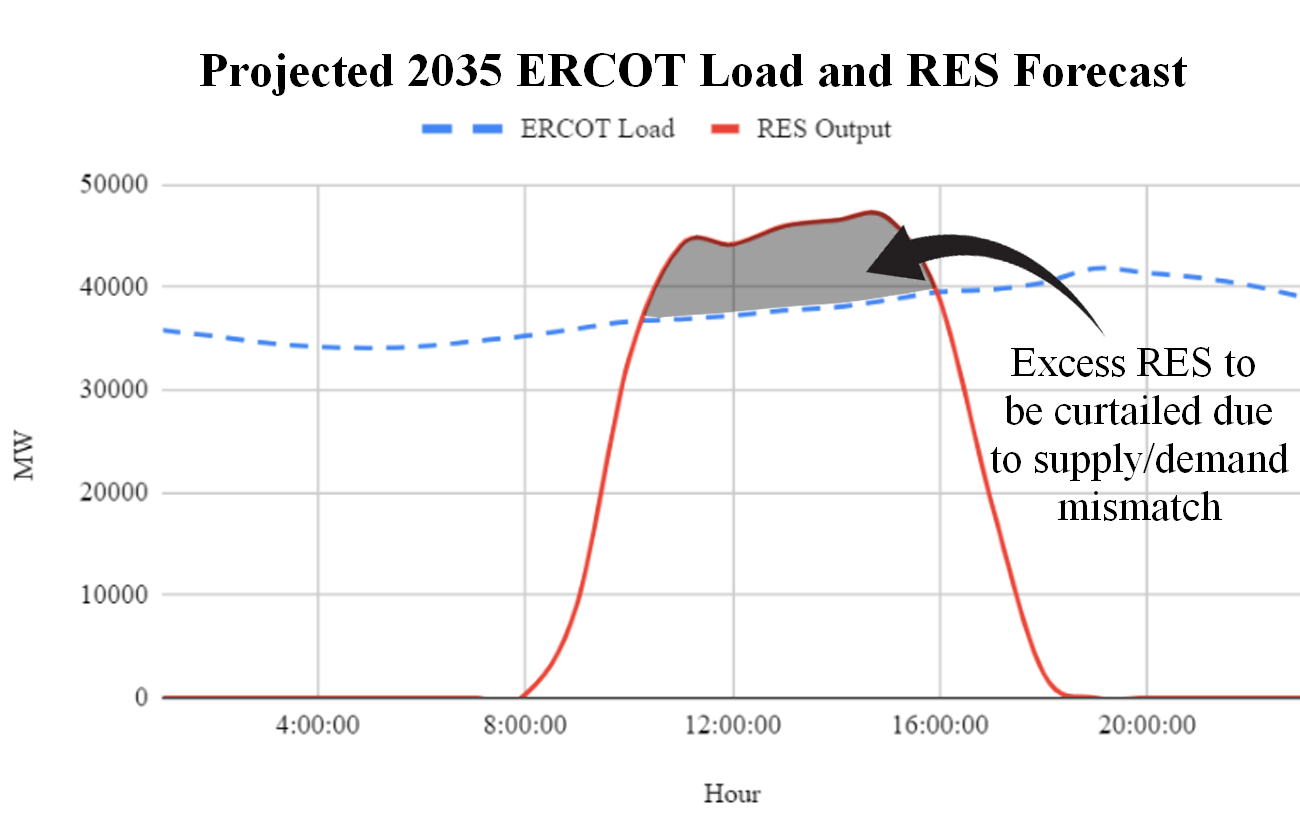}
\caption{When RES generation exceeds load, this can lead to cost effective, carbon free electricity sources such as wind and solar going underutilized. The underutilized RES output is the highlighted region, showing the difference between output and load.}
\label{RESCurtailmentExample}
\end{figure}

Meanwhile, the Electric Reliability Council of Texas (ERCOT)---the grid that services 90\% of Texas residents---has plans to substantially expand its renewable energy source (RES) capacity with several wind and solar projects by the year 2035 \cite{EIA}. These additions mean that if load profiles do not shift, there will likely be times where RES capacity alone exceeds total demand within this grid, which would result in the undesirable curtailment of these cost-effective, carbon-free electricity sources, as shown in Figure \ref{RESCurtailmentExample} (source profiles are obtained using the data and methods in Section \ref{Data and Methods}).

As power systems trend towards greater utilization of renewables due to their increased affordability and low emissions, they are also experiencing a trend towards increased connectivity \cite{van2021beyond, rimal2022smart}. For EVs, this connectivity comes in the form of smart electric vehicle supply equipment (EVSE) stemming from the growing internet-of-things (IoT) ecosystem. EVSE includes EV charging infrastructure that can receive direct communication such as demand response messages or price signals. Such communication allows for these aggregators to leverage enough EV charging profiles to influence and shift loads \cite{DOE2016, mohammadi2023towards, MohammadBOOKCHAPTER}. 

\begin{figure}[htbp]
\centering
\includegraphics[scale = .75]{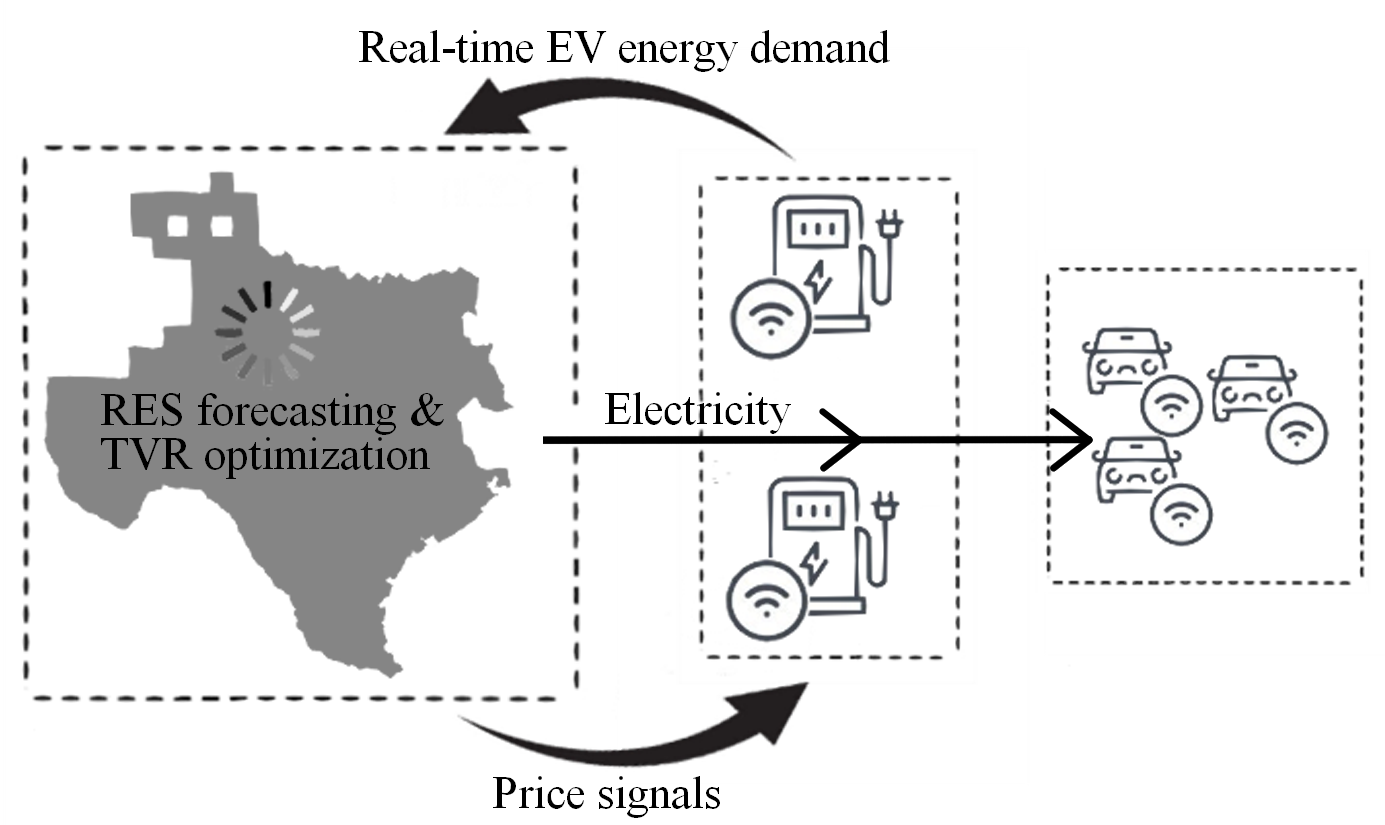}
\caption{Grids such as ERCOT can use RES forecasting and price signal based control schemes to manage their EVs through communication with chargepoint operators}
\label{MPCSummary}
\end{figure}

The ability to shift EV loads by leveraging IoT connectivity provides a valuable opportunity for grid operators to limit RES curtailment. For instance, real-time demand and generation information enables implementation of time-varying rates (TVRs). These TVRs incentivize or discourage EV charging depending on the needs of the grid by changing the price of electricity for EV chargers throughout the day. Figure \ref{MPCSummary} shows how, from a high level, a grid such as ERCOT could leverage its forecasting resources to coordinate with EV owners. This coordination is facilitated through smart charging equipment that is integrated within the IoT ecosystem and capable of receiving price signals and relaying real time data back to ERCOT about EV electricity usage.

While TVRs have been implemented by various utilities across the United States, they are not based on real-time inputs and therefore not updated regularly to account for fuel mix changes or EV charging demand. TVRs neglect this information because prior to the integration of a substantial amount of weather-dependent generation sources, TVRs were most commonly used to limit a relatively predictable peak demand. However, as grids such as ERCOT evolve to include highly variable wind and solar generation, and as EV penetration grows, it is important to consider different price structures that will best take advantage of RES output.

\begin{figure}[htbp]
\centering
\includegraphics[scale=.25]{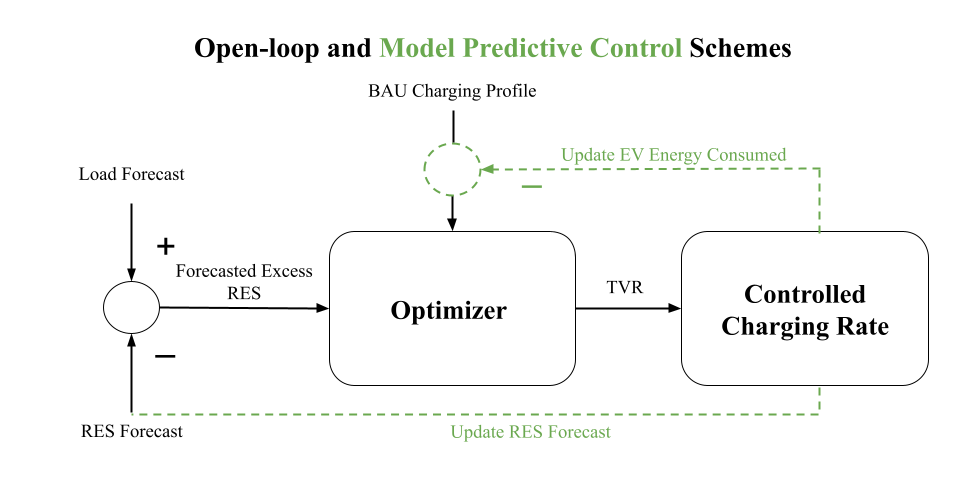}
\caption{Block diagrams for open-loop and model predictive control for the light duty EV fleet to harness excess renewable energy sources (RES) through price signaling with time varying rates (TVRs) given a business-as-usual (BAU) charging profile. Differences between open-loop and model predictive control (MPC) are shown in green dashed lines and inputs.}
\label{EV Control Schemes}
\end{figure}

One potential way to address the variability of RES is to use an optimization algorithm which takes into account day-ahead forecasting for RES output and grid-wide demand to create a TVR price structure that will shift EV load towards excess RES output. A simple way to implement this day-ahead forecast is shown in black solid lines in Figure \ref{EV Control Schemes}, which illustrates an open-loop control model where the input variable of the TVR shifts EV loads as they respond to these price signals. This open-loop control structure, however, would not be able to respond to any deviations in RES output from the prior day's forecast due to a lack of feedback mechanism for reference tracking. 

To account for this limitation of open-loop control, this study also investigates the benefits of using a model predictive control (MPC) scheme, where the same optimizer is executed for the next 24-hour forecast horizon, but the resulting pricing strategy is updated only for the next time step ahead. As the time horizon recedes with each update pricing strategy update, the RES forecast and electricity consumption of the light duty vehicle (LDV) fleet is also updated. This process is illustrated as a block diagram in Figure \ref{EV Control Schemes}.

To evaluate the effectiveness of open-loop and MPC control to inform price signaling and thereby limit RES curtailment by shifting EV charging, RES curtailment under the following control scenarios are considered: (1) RES curtailment from open-loop control price signaling, (2) RES curtailment using MPC to create and send price signals in 6 hour time steps, and (3) RES curtailment using MPC to create and send price signals in 3 hour time steps. Each control scheme's effectiveness is quantified in terms of the RES curtailment it avoids when compared to uncontrolled, "business as usual (BAU)" EV charging patterns.   

\subsection{Literature Review}


This section highlights works within the existing body of literature that provide important foundational insights regarding flexible loads, demand response, and control schemes within power systems. It also describes how they informed the scope of this study while highlighting the research gaps that we fill with our work's contributions. 

Authors in \cite{DONGOL} address the residential load mismatch between electricity demand and electricity production given the increasing amount of households with solar PV connected to the grid \cite{DONGOL}. The authors demonstrate that Model Predictive Control (MPC) can be applied to an existing grid architecture to maximize battery usage while minimizing PV energy supply peaks and residential energy demand through the addition of minimal hardware and software interfaces \cite{DONGOL}. However, the authors do contend that MPC has limitations for real time applications \cite{DONGOL}, meaning that balancing optimal control with realistic deployment is imperative.  

The researchers in \cite{Wang} analyze vehicle-to-grid (V2G) systems for the purpose of peak shaving and valley filling by trying to match a control target power curve with a planned V2G curve \cite{Wang}. They find that if the population of EVs capable of V2G is large enough in any one area, then their V2G control algorithm can be used singularly for peak shaving and valley filling \cite{Wang}. Although they are looking at a slightly different problem where V2G is accessible, the authors show that a residential EV fleet is capable of supporting the grid with valuable load-smoothing capabilities.   

In \cite{Diaz}, MPC was applied to a day-ahead scheduling optimization problem for charging an EV fleet \cite{Diaz}. They compare their results using MPC with an open-loop strategy and find that even though the MPC system is able to better compensate for generation disturbances and deviations in the EVs' starting level of charge, allowing for better fulfillment of the desired departure level of charge, the open-loop control methodology performed well \cite{Diaz}. Although less nuanced than the MPC, the open-loop control provides a more accessible control algorithm for residential users who do not have the means or desire to coordinate a precise charging algorithm. \cite{Diaz} validates the use of MPC as a control mechanism for EV fleets, which our work extends to grids with highly variable RES outputs to further investigate its benefits. 

Authors in \cite{CEDILLO} conduct simulations based on a bi-level optimization approach and a stochastic EV charging bid and EV charging schedule control strategy to maximize alternative revenue streams \cite{CEDILLO}. Overall, they show that their proposed dynamic pricing strategies do influence EV owners to charge or discharge in such a way as to provide grid balancing services \cite{CEDILLO}. They suggest further areas of research to include exploration of demand curves to look at rates for different customer groups given a more elastic or inelastic demand response \cite{CEDILLO}. This study's demonstration of EV owner participation in demand response measures is promising for research looking at how to leverage EVs as a flexible load.

In their study, the authors of \cite{Khalid} use a Game Theory-based TVR model for on-peak and shoulder-peak hour pricing \cite{Khalid}. Of note, the authors find that shifting load to off-peak hours from peak hours may cause ``rebound peaks" whereby consumers subject to the same pricing mechanism increase the demand on the grid disproportionately following the peak pricing \cite{Khalid}. Depending on the grid and RES profiles, this rebound effect could have negative impacts on the goal of limiting RES curtailment. This finding is the reason that MPC in different time steps is examined in this work, as longer time steps could mitigate the oscillatory aggregated consumer behavior that can lead to rebound peaks from such frequent potential over-corrections. 

The solutions presented in this literature, while foundational for understanding the growing role of EVs within power systems, could benefit from the further consideration of other evolving factors within power systems. Such factors include the growing, flexible EV load, increased intermittent RES penetration levels, and the increased interconnectivity of power systems as the IoT ecosystem evolves. For these reasons, our study focuses on control mechanisms which leverage communication between EV chargers and the grid to minimize RES curtailment, which will continue becoming an increasingly important consideration for controlling both the cost and emissions of the grid. 

\subsection{Contributions and Paper Structure}

This work contributes a case study built upon real-world data and realistically-derived parameters to demonstrate how EVs can support the transition of power grids as they become more distributed and renewable. Two feasible time-varying rate (TVR) control schemes (open loop and model predictive control) are evaluated. Section \ref{Data and Methods} outlines the data used to source or derive inputs for the control schemes studied in this work and the control methodologies used. Next, Section \ref{Results and Discussion} presents the results of using the control schemes evaluated in this paper when compared to a business as usual scenario where no control is used. Trends found within the results and interpretations for their real world implications are provided. Lastly, Section \ref{Conclusions} outlines pertinent takeaways from the results and outlines directions for future works. 

\section{Methods and Data}
\label{Data and Methods}

\begin{figure}[htbp]
\centering
\includegraphics[scale = .7]{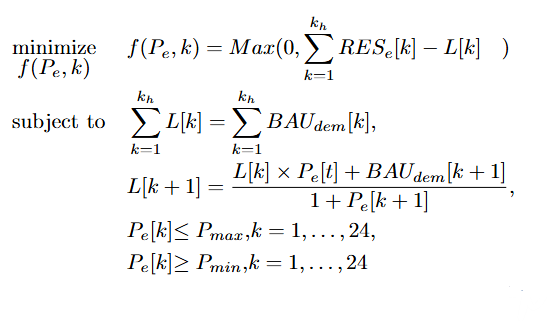}
\caption{The formulation of the optimization problem used in this study's control schemes.}
\label{optimizer}
\end{figure}

For all control schemes examined, this case study uses an optimizer to find a time-varying rate (TVR) profile that results in the lowest RES curtailment during a given day. For open-loop control this rate profile is calculated using a day-ahead forecast and then implemented at the start of the following day. For MPC, a 3- or 6-hour time step is employed. At the conclusion of each time step, the optimizer updates its forecast to account for changes in RES output projections and consumed EV energy. All schemes are optimized to minimize RES curtailment and use constraints that maintain total EV energy demand equivalent to the business as usual (BAU) energy demand throughout a 24-hour period, meaning that EV load is redistributed (e.g., load shifting) in accordance with the effectiveness of the TVR employed. Block diagrams of these control schemes are shown in Figure \ref{EV Control Schemes} and the optimization formulation is shown in Figure \ref{optimizer}. 

The objective function to be minimized is the sum of the excess wind and solar renewable energy sources (RES) over each hour $k$, ($RES_e[k]$) that would be curtailed ($\max(RES_e[k]$ - $L[k], 0)$). This objective function is subject to the summation of the aggregated EV load, $L$ being equal to the summation of the uncontrolled, BAU charging profile, $BAU_{dem}[k]$, to ensure that total EV charging demands are still being met each day. This objective is summed over the time horizon, $k_h$, which is 24 hours for the open-loop control problem. For MPC, $k_h$ starts at 24 hours and then decreases by the value of each time step after each iteration. The first constraint describes the energy consumption of the EV fleet while the second constraint describes rate of energy consumption as a result of the price signaling of TVRs. $P_{e}$ is the percent of the load shifted from each hour $k$ by the TVR, and ${BAU_{dem}}$ is the BAU demand from the EV fleet. Lastly, $P_{e}$ is constrained by the percentage of EV owners that would modify their charging due to the price of the TVR. 

Under MPC, after each optimization time step, the energy consumed by the EV fleet within the time step is subtracted and redistributed across ${BAU_{dem}}$, and the next 3 hours of $RES[k]$ are updated to be the day's actual RES output for both MPC schemes. These updates are illustrated in the highlighted portions of Figure \ref{EV Control Schemes}.

The simulation of these control schemes takes place under expected conditions for ERCOT in the year 2035, where EV presence will be high enough for load shifting to yield significant changes and there will be enough RES capacity to generate output that exceeds grid-wide load \cite{EIAProjections}. The inputs required to run the simulations are (1) EV fleet conditions (EVs on the road in 2035, EV electricity use, and limitations for how much of the EV fleet we can expect will be willing to shift their load in response to price signals) and (2) grid conditions (a scaling factor for current grid wide loads to 2035 grid wide load, RES installed capacity in 2035, and profiles for day-ahead forecasted excess and day-of actual excess RES).

\subsection{EV Fleet Conditions}
This case study assumes that benchmarks for EV adoption within the United States that have been set in place by the Biden Administration are achieved in 2030 \cite{EVTarget}. They are then extrapolated beyond 2030 by using projections found in \cite{USDrive} for their ``high" EV fleet growth scenario. Market share benchmarks are modeled as being approached linearly over time, and the total number of registered EVs at the end of year $n$ is determined using Equation \eqref{EV_total_N}, where $EVt$ is the total number of EVs on the road, $MS$ is EV market share or percentage of light duty vehicles (LDVs) purchased in a given year that are electric, ${R}$ is the lifetime in years of the average vehicle, and ${LDV}$ is the total number of registered LDVs on the road. ${R}$ accounts for EVs being retired over time \cite{Keith_2019} and approximates the number of new vehicles being purchased to replace disposed vehicles each year.

\begin{equation}\label{EV_total_N}
EVt_{n}=\frac{R-1}{R}{EVt_{n-1}}+MS_{n}*\frac{LDV{n}}{R}
\end{equation}

The EV percentage, or percentage of LDVs on the road that are electric, from this adoption rate is then applied to the daily distance traveled within ERCOT serviced counties to give the total daily distance traveled by the EV fleet \cite{BTS2022}. These distances are then converted into a daily electricity demand (in $kWh$) from the average energy demand for electric LDVs (in $kWh/km$) \cite{Mileage2}. In order to distribute the load from these daily demands across an hourly profile, data from \cite{GridPIQ} was used, which provides average distributions for EV charging profiles by hour. These distributions were measured from several charging stations and households across Texas and are presented on a per-vehicle basis, meaning that they are scalable to EV fleet size.

Due to limitations in existing research on how time-varying rates can precisely be expected to influence EV charging behavior, the value for the maximum effectiveness of a TVR to delay charging, $P_{max}$, is approximated as 50\% in accordance with ranges provided in publications from the Department of Energy \cite{DOE2016} and the Rocky Mountain Institute \cite{RockyMountainInstitute}. 

\subsection{Grid Conditions}

\begin{figure}[htbp]
\centering
\includegraphics[scale = .25]{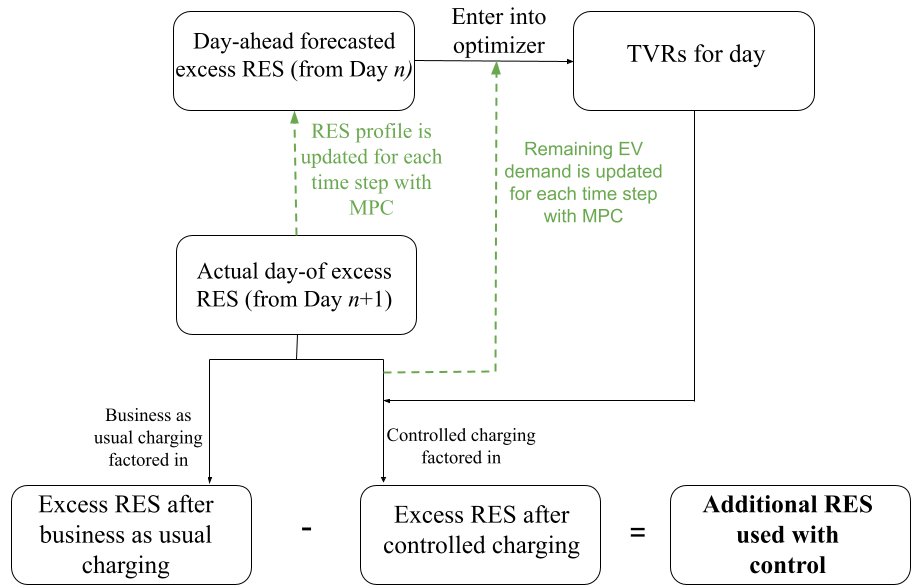}
\caption{TVRs are created and tested on two back to back RES profiles. The differences between open-loop and model predictive control (MPC) are shown in green dashed lines and inputs.}
\label{DataAndMethodsFigure}
\end{figure}

Projections from the U.S Energy Information Administration (EIA) were used for Texas's solar and wind installations \cite{EIAProjections}. Historical data from ERCOT's 2023 solar and wind output as a percentage of installed capacity were applied to these projected 2035 installations \cite{ERCOT}. Next, in order to account for population growth over the next decade, ERCOT demand was scaled to 2035 population projections from the Texas Demographic Center \cite{Population}. The combined loads and RES output were used to estimate remaining excess RES profiles that the EV loads can be shifted towards in order to limit curtailment. To simulate a seasonably-appropriate RES output that deviates from each of these day-ahead forecasts, the next excess RES profile within the year of data was used as the day-of, actual RES output. This actual output was used to update each MPC's optimizer at each time step. It was also used to evaluate the effectiveness of all the control schemes' TVRs. This process is shown in Figure \ref{DataAndMethodsFigure}.

\section{Results and Discussion}
\label{Results and Discussion}

Table \ref{Table1} shows the input values or source profiles for each of the defined model inputs from Section \ref{Data and Methods}.

\begin{table}[htbp]
\caption{2035 parameters for MPC model inputs} 
\label{Table1}
\begin{center}
\begin{tabular}{|c|c|c|}
\hline
\textbf{Inputs} & \textbf{Value or Source Profile}\\
 \hline
EVs & 9450000 LDVs \\
 \hline
MWh/day from EVs & 82443 MWh \\
 \hline
TX Population Increase & 13\% \cite{Population}\\
 \hline
BAU Profile & GridPIQ \cite{GridPIQ} \\
 \hline
RES Output & ERCOT \cite{ERCOT2} \& EIA \cite{EIAProjections} \\
\hline
TVR Uptake Percent & 50\% \cite{DOETVRs} \\
\hline

 \end{tabular}
 \end{center}
\end{table}

Inputs from Table \ref{Table1} were used in the optimization problem formulated in Equation \eqref{optimizer} and implemented across every instance of expected RES curtailment over one year of historical ERCOT data scaled to 2035 load, EV levels, and RES output. This year of scaled data ended up having 133 days of expected RES curtailment. For the remaining days, ERCOT load was never higher than RES output, so there was no additional RES supply to try to manage through demand response measures. Each of the three control schemes investigated in this paper (open-loop control, MPC with 6 hour time steps, and MPC with 3 hour time steps), were simulated for these 133 days to see how much additional RES could be utilized by shifting EV loads. The results of each control scheme over the course of the entire year are summarized in Table \ref{YearSummary}.

\begin{table}[htbp]
\caption{One year summary of RES curtailment when compared to no control being used.} 
\label{YearSummary}
\begin{center}
\begin{tabular}{|p{3.5cm}|p{.9cm}|p{.9cm}|p{.9cm}|}
\hline
 & Open-Loop & MPC 6 Hours & MPC 3 Hours\\
 \hline
Increased RES utilized over 1 year when compared with no control & 165 GWh & 518 GWh & 818 GWh \\
 \hline
Frequency of days this control scheme utilized the most RES & 0\% & 29\% & 71\% \\
 \hline
\# Of days this scheme \textbf{decreased} RES utilized when compared with no control & 13 days & 0 days & 4 days \\
 \hline


 \end{tabular}
 \end{center}
\end{table}

The first row of Table \ref{YearSummary} shows the quantity of additional RES utilized, in GWh, over the course of a year under each control scheme when compared to a baseline scenario with no control. The results show that over the course of a year, MPC with a smaller time step of 3 hours, cumulatively, utilizes the most RES out of the 3 control schemes. The second row shows the percentage of days out of the year that each control scheme prevailed by resulting in the greatest decrease in RES curtailment of the three options. Lastly, the third row shows the number of days each scheme resulted in \textit{increased} RES curtailment (or an overall decrease in RES use) when compared to business as usual (uncontrolled) charging. This poor performance can likely be attributed to forecast inaccuracies or price signals that were too frequent and resulted in oscillatory charging behavior \cite{SummerRebound}. It is noteworthy that MPC in 6 hour time steps avoided all instances of being worse for the day's RES usage than no control at all. 

For a more granular view of the control schemes, the results from the first sample day of each month are presented in Table \ref{SampleTable}. These results are presented as additional RES utilized in MWh when compared to the business as usual case where no control is used. Each sample day's highest amount of additional RES used is shown in \textbf{bold}.

\begin{table}[htbp]
\caption{Additional RES utilized for the first sample day from each month by control scheme} 
\label{SampleTable}
\begin{center}
\begin{tabular}{|c|c|c|c|}
\hline
&
\multicolumn{3}{|c|}{Additional RES used (MWh) by control scheme} \\
\hline
Sample Day & Open-Loop & MPC 6 Hours & MPC 3 Hours\\
 \hline
January & 567 & 567 & \textbf{1170}\\
 \hline
February & 659 & 6323 & \textbf{16123}\\
 \hline
March & 9661 & \textbf{11781} & 8867\\
 \hline
April & 166 & \textbf{3755} & 1872\\
 \hline

May & -287 & \textbf{3114} & 1763\\
 \hline

June & 328 & \textbf{2516} & 395\\
 \hline

July & -1.3 & 1134 & \textbf{1390}\\
 \hline

August & 121 & 2018 & \textbf{2968}\\
 \hline

September & 490 & 1553 & \textbf{2649}\\
 \hline

October & 778 & 6185 & \textbf{11658}\\
 \hline

November & 4768 & 6747 & \textbf{10093}\\
 \hline

December & 278 & 4152 & \textbf{5520}\\
 \hline
 \end{tabular}
 \end{center}
\end{table}

As discussed earlier in reference to Table \ref{YearSummary}, the negative values that occur in Table \ref{SampleTable} show that less RES is utilized on these days than under the uncontrolled case where no price signaling information is used at all. This poor performance occurs because open-loop control lacks a feedback mechanism to account for forecast inaccuracies. MPC in 3 hour time steps results in greater utilization of RES for the largest number of days due to its ability to correct for forecast inaccuracies more frequently than MPC in 6 hour time steps. MPC in 6 hour time steps, however, sometimes outperforms MPC in 3 hour time steps. This outcome may be attributed the fact that attempting to influence charging behaviour too frequently has been shown to lead to oscillatory behaviour, sometimes even creating secondary, rebound peaks \cite{SummerRebound}. 

\begin{figure}[htbp]
\centering
\includegraphics[scale=.45]{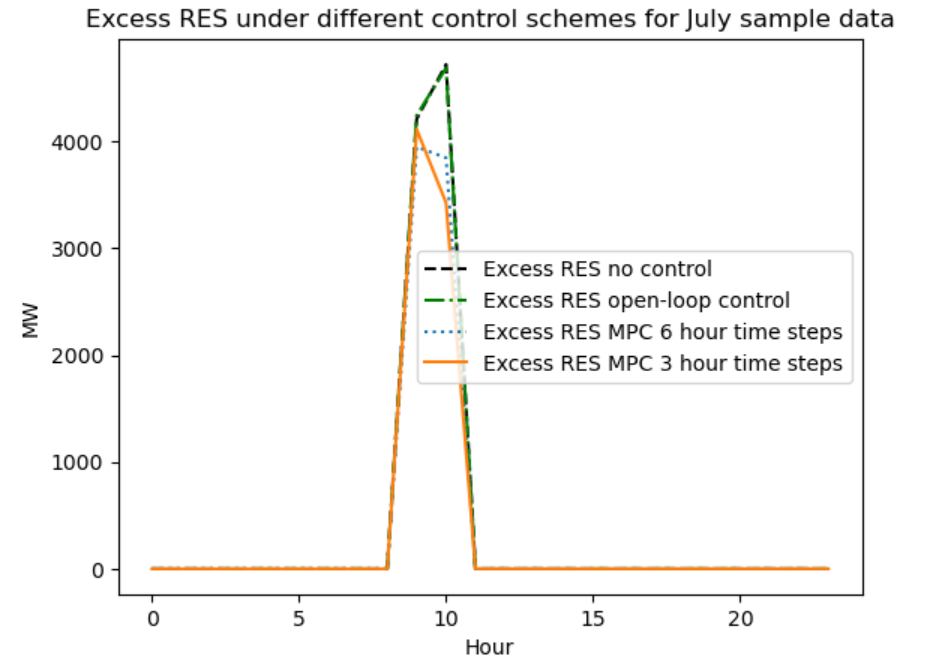}
\caption{The area under each curve shows how much RES went unutilized at the end of the sample day. Smaller areas indicate greater RES use.}
\label{RESExample}
\end{figure}

Figure \ref{RESExample} graphically illustrates remaining excess RES (or RES that would be curtailed and therefore go unutilized) during a sample day in July for each control scheme. The graph shows that the BAU case with no control has the greatest level of excess RES, indicated by the greatest area underneath the curve (8933 MWh). Open-loop control has a similar level of excess RES (8934 MWh), but MPC with 6-hour and 3-hour time steps decrease remaining excess RES (7798 Mwh and 7542MWh respectively) .

\section{Conclusions}
\label{Conclusions}

This case study demonstrates that while day-ahead forecasting can improve the average performance of time-varying rates to harness RES supply in a high-renewable grid, model predictive control further increases the amount of RES utilized by responding to forecast deviations. The results also highlight that in some cases, updating rates too frequently may have the opposite intended effect through the creation of rebound peaks from oscillatory charging behavior. Grid operators should consider balancing the benefits of using more RES with mitigating downsides of charging control, such as rebound peaks. Operators should also select a control scheme that provides enough time for data curation, and to avoid overwhelming EV users with frequent rate updates. The results of this study show that MPC in 6 hour time steps is a viable option for grid operators looking to balance these benefits and downsides of price signaling to EV owners. 

Future works should expand upon this research by investigating the effect of varying on/off peak price ratios on the desired $P_e$, or TVR uptake percentage. Future studies should also examine the frequency and magnitude by which RES output deviates from forecasts, and how these deviations affect the performance of pricing control schemes. Presently, ERCOT's website states that ``the [forecasting] data in this graph should not be used to evaluate forecast performance" \cite{ERCOTForecasting}, but if forecasting data suitable for evaluating performance becomes available through ERCOT it can be incorporated into future frameworks. Similarly, different scenarios for residential load projections should be considered. Lastly, future works should look at leveraging electric vehicles outside of the LDV fleet, such as electric freight or aviation as these technologies mature in the coming years. 

\bibliographystyle{IEEEtran}
\bibliography{main.bib}

\begin{thebibliography}{10}
\providecommand{\url}[1]{#1}
\csname url@samestyle\endcsname
\providecommand{\newblock}{\relax}
\providecommand{\bibinfo}[2]{#2}
\providecommand{\BIBentrySTDinterwordspacing}{\spaceskip=0pt\relax}
\providecommand{\BIBentryALTinterwordstretchfactor}{4}
\providecommand{\BIBentryALTinterwordspacing}{\spaceskip=\fontdimen2\font plus
\BIBentryALTinterwordstretchfactor\fontdimen3\font minus \fontdimen4\font\relax}
\providecommand{\BIBforeignlanguage}[2]{{%
\expandafter\ifx\csname l@#1\endcsname\relax
\typeout{** WARNING: IEEEtran.bst: No hyphenation pattern has been}%
\typeout{** loaded for the language `#1'. Using the pattern for}%
\typeout{** the default language instead.}%
\else
\language=\csname l@#1\endcsname
\fi
#2}}
\providecommand{\BIBdecl}{\relax}
\BIBdecl

\bibitem{Registrations}
Ruder, Adam. “State EV Registration Data.” Atlas EV Hub, 7 Sept. 2022.

\bibitem{EVTarget}
(2021, December) Fact sheet: The biden-harris electric vehicle charging action plan. The e. Washington, D.C.

\bibitem{Keith_2019}
\BIBentryALTinterwordspacing
D.~R. Keith, S.~Houston, and S.~Naumov, ``Vehicle fleet turnover and the future of fuel economy,'' \emph{Environmental Research Letters}, vol.~14, no.~2, p. 021001, feb 2019. [Online]. Available: \url{https://dx.doi.org/10.1088/1748-9326/aaf4d2}
\BIBentrySTDinterwordspacing

\bibitem{EIA}
\BIBentryALTinterwordspacing
U.~E.~I. Administration. Electricity explained: Factors affecting electricity prices. [Online]. Available: \url{https://www.eia.gov/energyexplained/electricity/prices-and-factors-affecting-prices.php}
\BIBentrySTDinterwordspacing

\bibitem{van2021beyond}
J.~Van~Mierlo, M.~Berecibar, M.~El~Baghdadi, C.~De~Cauwer, M.~Messagie, T.~Coosemans, V.~A. Jacobs, and O.~Hegazy, ``Beyond the state of the art of electric vehicles: A fact-based paper of the current and prospective electric vehicle technologies,'' \emph{World Electric Vehicle Journal}, vol.~12, no.~1, p.~20, 2021.

\bibitem{rimal2022smart}
B.~P. Rimal, C.~Kong, B.~Poudel, Y.~Wang, and P.~Shahi, ``Smart electric vehicle charging in the era of internet of vehicles, emerging trends, and open issues,'' \emph{Energies}, vol.~15, no.~5, p. 1908, 2022.

\bibitem{DOE2016}
\BIBentryALTinterwordspacing
Customer acceptance, retention, and response to time-based rates from the consumer behavior studies. [Online]. Available: \url{https://www.energy.gov/sites/prod/files/2016/12/f34/CBS_Final_Program_Impact_Report_Draft_20161101_0.pdf}
\BIBentrySTDinterwordspacing

\bibitem{mohammadi2023towards}
M.~Mohammadi, J.~Thornburg, and J.~Mohammadi, ``Towards an energy future with ubiquitous electric vehicles: Barriers and opportunities,'' \emph{Energies}, vol.~16, no.~17, p. 6379, 2023.

\bibitem{MohammadBOOKCHAPTER}
M.~Mohammadi and A.~Mohammadi, ``Empowering distributed solutions in renewable energy systems and grid optimization,'' pp. 1--17.

\bibitem{DONGOL}
\BIBentryALTinterwordspacing
D.~Dongol, T.~Feldmann, M.~Schmidt, and E.~Bollin, ``A model predictive control based peak shaving application of battery for a household with photovoltaic system in a rural distribution grid,'' \emph{Sustainable Energy, Grids and Networks}, vol.~16, pp. 1--13, 2018. [Online]. Available: \url{https://www.sciencedirect.com/science/article/pii/S2352467717302916}
\BIBentrySTDinterwordspacing

\bibitem{Wang}
Z.~Wang and S.~Wang, ``Grid power peak shaving and valley filling using vehicle-to-grid systems,'' \emph{IEEE Transactions on Power Delivery}, vol.~28, no.~3, pp. 1822--1829, 2013.

\bibitem{Diaz}
C.~Diaz, A.~Mazza, F.~Ruiz, D.~Patino, and G.~Chicco, ``Understanding model predictive control for electric vehicle charging dispatch,'' in \emph{2018 53rd International Universities Power Engineering Conference (UPEC)}, 2018, pp. 1--6.

\bibitem{CEDILLO}
\BIBentryALTinterwordspacing
M.~H. Cedillo, H.~Sun, J.~Jiang, and Y.~Cao, ``Dynamic pricing and control for ev charging stations with solar generation,'' \emph{Applied Energy}, vol. 326, p. 119920, 2022. [Online]. Available: \url{https://www.sciencedirect.com/science/article/pii/S0306261922011771}
\BIBentrySTDinterwordspacing

\bibitem{Khalid}
\BIBentryALTinterwordspacing
A.~Khalid, N.~Javaid, A.~Mateen, M.~Ilahi, T.~Saba, and A.~Rehman, ``Enhanced time-of-use electricity price rate using game theory,'' \emph{Electronics}, vol.~8, no.~1, 2019. [Online]. Available: \url{https://www.mdpi.com/2079-9292/8/1/48}
\BIBentrySTDinterwordspacing

\bibitem{EIAProjections}
\BIBentryALTinterwordspacing
(2023) As texas wind and solar capacity increase, energy curtailments are also likely to rise. [Online]. Available: \url{https://www.eia.gov/todayinenergy/detail.php?id=57100}
\BIBentrySTDinterwordspacing

\bibitem{USDrive}
{Grid Integration Tech Team and Integrated Systems Analysis Tech Team}, ``Summary report on evs at scale and the u.s. electric power system,'' US Drive, Tech. Rep. 3002011613, November 2019.

\bibitem{BTS2022}
\BIBentryALTinterwordspacing
M.~Liu. (2022, July) Trips by distance. Bureau of Transportation Statistics. [Online]. Available: \url{https://data.bts.gov/Research-and-Statistics/Trips-by-Distance/w96p-f2qv}
\BIBentrySTDinterwordspacing

\bibitem{Mileage2}
\BIBentryALTinterwordspacing
A.~V.~T. Laboratory. Comparing energy costs per mile for electric and gasoline-fueled vehicles. Idaho National Laboratory. Washington, D.C. [Online]. Available: \url{https://avt.inl.gov/sites/default/files/pdf/fsev/costs.pdf}
\BIBentrySTDinterwordspacing

\bibitem{GridPIQ}
\BIBentryALTinterwordspacing
GridPIQ. (2021, August) Uncoordinated charging profile. Pacific Northwest National Laboratory. Richland, Washington. [Online]. Available: \url{https://gridpiq.pnnl.gov}
\BIBentrySTDinterwordspacing

\bibitem{RockyMountainInstitute}
\BIBentryALTinterwordspacing
R.~M. Institute. (2016) A review of alernative rate designs. [Online]. Available: \url{https://rmi.org/wp-content/uploads/2017/04/A-Review-of-Alternative-Rate-Designs-2016.pdf}
\BIBentrySTDinterwordspacing

\bibitem{ERCOT}
\BIBentryALTinterwordspacing
E.~R.~C. of~Texas. (2023) Hourly load data archives. [Online]. Available: \url{https://www.ercot.com/gridinfo/load/load_hist}
\BIBentrySTDinterwordspacing

\bibitem{Population}
\BIBentryALTinterwordspacing
H.~You. (2023) Estimates and projections for texas. Texas Demographic Center Population Estimates Program. [Online]. Available: \url{https://demographics.texas.gov/Resources/Presentations/DDUC/2023/2023_05_24_PopulationEstimatesProjectionsTexas.pdf}
\BIBentrySTDinterwordspacing

\bibitem{ERCOT2}
\BIBentryALTinterwordspacing
(2022) Wind and solar integration reports. [Online]. Available: \url{https://www.ercot.com/gridinfo/generation/windandsolar}
\BIBentrySTDinterwordspacing

\bibitem{DOETVRs}
S.~G. I.~G. Program, ``Customer acceptance, retention, and response to time-based rates from the consumer behavior studies,'' US Department of Energy, Tech. Rep., 2016.

\bibitem{SummerRebound}
\BIBentryALTinterwordspacing
Y.~L. Qiu, Y.~D. Wang, H.~Iseki, X.~Shen, B.~Xing, and H.~Zhang, ``Empirical grid impact of in-home electric vehicle charging differs from predictions,'' \emph{Resource and Energy Economics}, vol.~67, p. 101275, 2022. [Online]. Available: \url{https://www.sciencedirect.com/science/article/pii/S0928765521000609}
\BIBentrySTDinterwordspacing

\bibitem{ERCOTForecasting}
\BIBentryALTinterwordspacing
(2024) Combined wind and solar. [Online]. Available: \url{https://www.ercot.com/gridmktinfo/dashboards/combinedwindandsolar}
\BIBentrySTDinterwordspacing

\end{thebibliography}

\end{document}